# The Coriolis Effect Further Described in the Seventeenth Century

Christopher M. Graney
(christopher.graney@kctcs.edu)
Jefferson Community & Technical College
Louisville, Kentucky (USA)

ABSTRACT: Claude Francis Milliet Dechales described the Coriolis effect in his 1674 *Cursus seu Mundus Mathematicus*. Dechales discussed and illustrated the deflection of both falling bodies and of projectiles launched toward the poles that should occur on a rotating Earth. Interestingly, this was done as an argument against the Earth's rotation, the deflections not having been observed at the time. Dechales's work follows on that of Giovanni Battista Riccioli, who had also described the effect in his *Almagestum Novum* of 1651.

On Earth's surface, projectiles and bodies in free fall appear to be deflected by the Earth's rotation. This is called the "Coriolis effect," after Gaspard-Gustave de Coriolis (1792-1843) who described it mathematically in 1835.[1] The Jesuit astronomer Giovanni Battista Riccioli (1598-1671), who studied falling bodies and produced the first precise measurements of *g*, foresaw the effect and described it in his 1651 *Almagestum Novum* as an argument against the motion of the Earth (the effect's non-detection being taken as indication of Earth's immobility). Riccioli's description seems to be the earliest description of the effect found so far (see Figure 1).[2]

Riccioli's work might be presumed to be a historical anomaly—an insight published once and forgotten. However, I have recently discovered another discussion of the Coriolis effect in a Jesuit work, again as an argument against the motion of the Earth.[3] (Note that anti-Copernicans such as Riccioli and Dechales held to a hybrid geocentric, or "geo-heliocentric", theory developed by Tycho Brahe—in which the sun, moon, and stars circled an immobile Earth while the planets circled the sun—a theory compatible with the telescopic discoveries of the seventeenth century.) This is in the 1674 *Cursus seu Mundus Mathematicus* of Claude Francis Milliet Dechales (1621-1678), republished in 1690.

In a section entitled "Objectiones contra Copernicum," Dechales notes that the "common" or "vulgar" objections to the Copernican motion of Earth all fail—for example, the objection that a rotating Earth will leave behind birds in flight. Dechales illustrates why using the example of a ball released from the yardarm of a ship (Figure 2): the ball is not left behind if the ship is in motion; rather, on account of common motion, if the ship is moving steadily then the ball falls to the same spot as it does when the ship is at rest.

But then Dechales includes a diagram (Figure 3) of a tower carried by a rotating Earth, and asks his reader to consider the following:

> *A ball F, hanging from the top of a tower directly above point G, is dropped. While the ball descends, point G is carried [by Earth's rotation] into I. I propose the ball F to be unable to arrive at point G (now at I). This is because the ball when positioned at F has a momentum [*impetus[4]*] requisite for passing through arc FH*

> *(through which the tower top moves while the ball descends) which is greater than that requisite for arc GI. Therefore, if the ball is dropped, it will not arrive at point I, but will advance forward farther [to L].*

Dechales also includes a diagram (Figure 4) of a cannon, and continues his discussion:

> *Likewise, consider a cannon discharged in the direction of one of the poles [of the Earth]. During the time in which the ball would traverse the distance MO, the cannon [originally at M] would be carried to N, while the target [originally at O] would be carried to P. While the ball was at M, it had from the rotation of the Earth momentum able to carry it through arc MN (which is greater than arc OP) during the time required to travel from M to O. But the ball, separated from the cannon, conserves this momentum whole [*totum hunc impetum conservat*]. Therefore it should run through a distance greater than OP, and consequently not hit the target.*

Dechales illustrates the ball passing to the right of the target. Dechales cites Riccioli repeatedly in the *Cursus*, but his diagrams differ from Riccioli's.

Thus Dechales illustrates how a rotating Earth should produce detectable effects in the motions of falling bodies and of projectiles.[5] Dechales's qualitative treatment of these effects differs very little from modern qualitative discussions of the Coriolis effect.[6] What we now call the "Coriolis" effect was being described and illustrated by different authors a century before Coriolis was born. Riccioli was not an anomaly.

A remarkable twist to this story is that the effect was first described via an anti-Copernican argument. Nonetheless, if we grant honor to "firsts" in science, it seems the "Coriolis" effect might be due for a renaming.

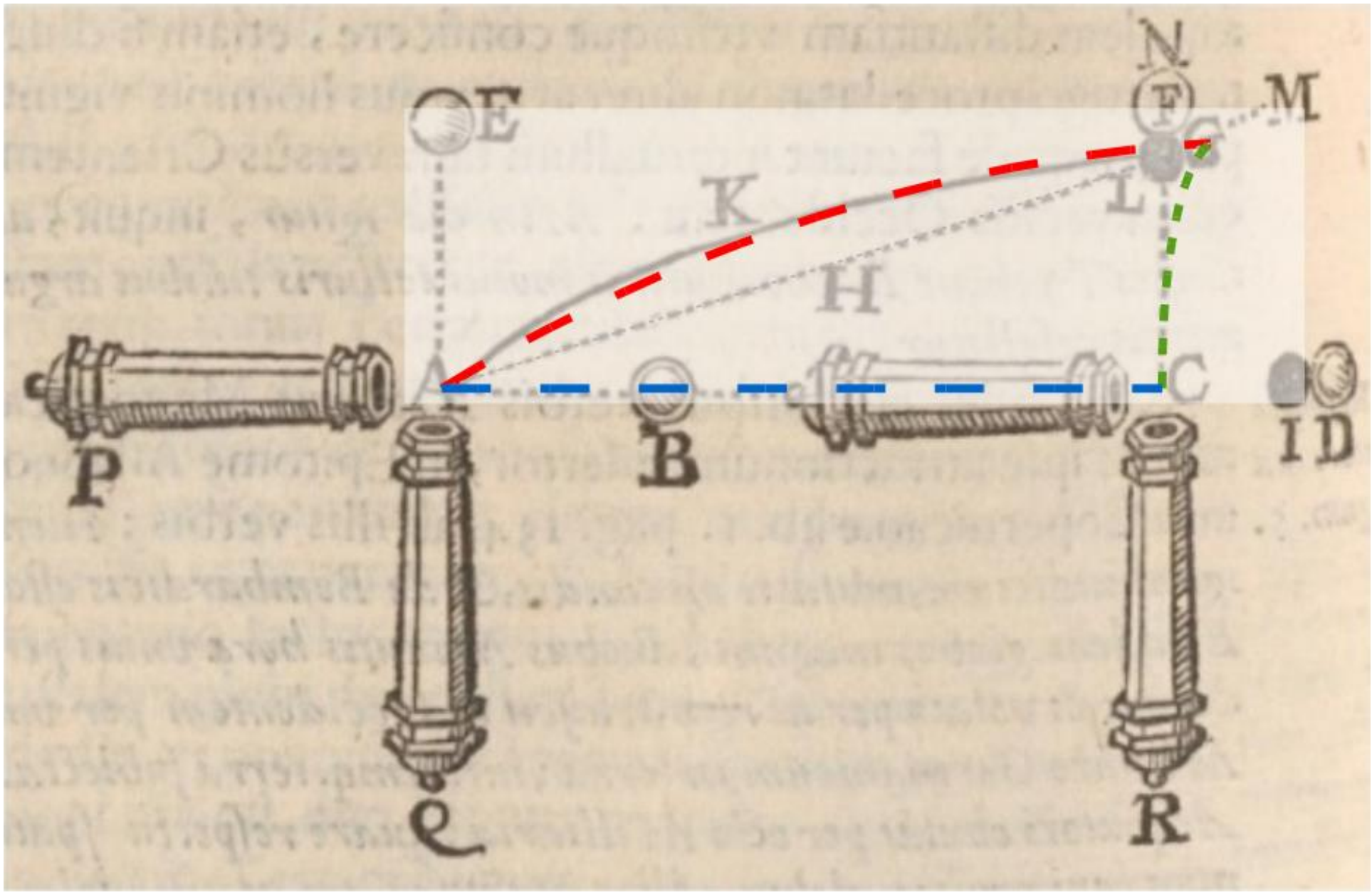

**Figure 1:** Illustration from Riccioli's *Almagestum Novum*, with superimposed plots from simple Coriolis effect calculations. A cannon with mouth at A is fired toward eastern target B and northern target E, both equally distant from A. While the cannon ball (I and F) is in flight, the diurnal rotation of the Earth carries the cannon mouth to C, the eastern target to D, and the northern target to N. Colored dashed lines show calculated paths for a cannon in a rotating frame of reference, firing a ball toward the axis of rotation: the blue line is the path of the cannon's mouth; the red line is the path of the ball; the green line is the path of the ball as seen from the cannon, drawn from position C. Up (north) is toward the axis of rotation. The scale of the plot has been fit to the illustration, and the distance to the axis of rotation has been adjusted so as to yield agreement between the red line and curve AKF. Nevertheless, note the rightward (eastward) deflection of the ball as seen from the cannon, away from line of fire CF and into G, which Riccioli describes.[7] All images credit: ETH-Bibliothek Zürich, Alte und Seltene Drucke.

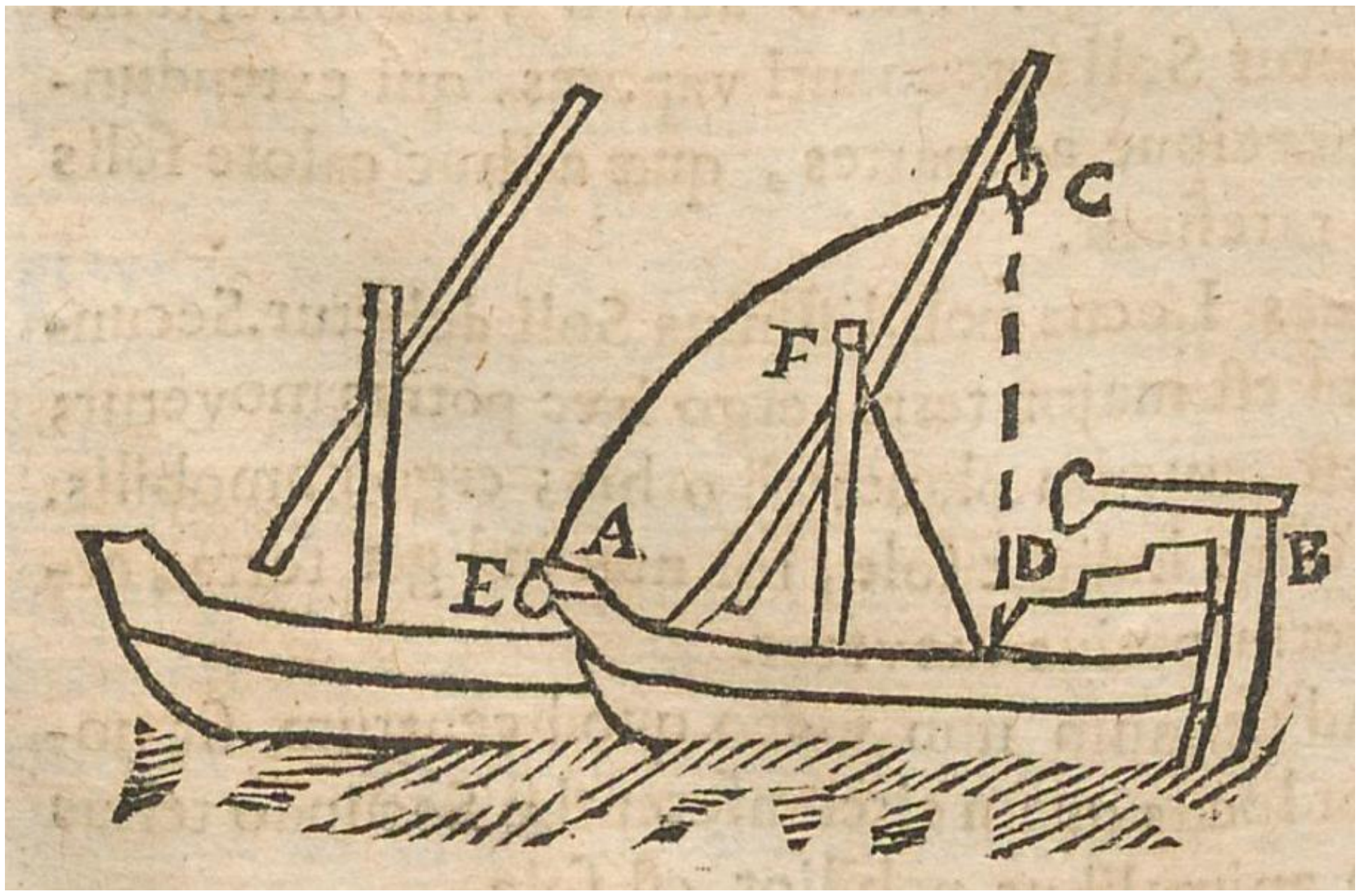

**Figure 2**

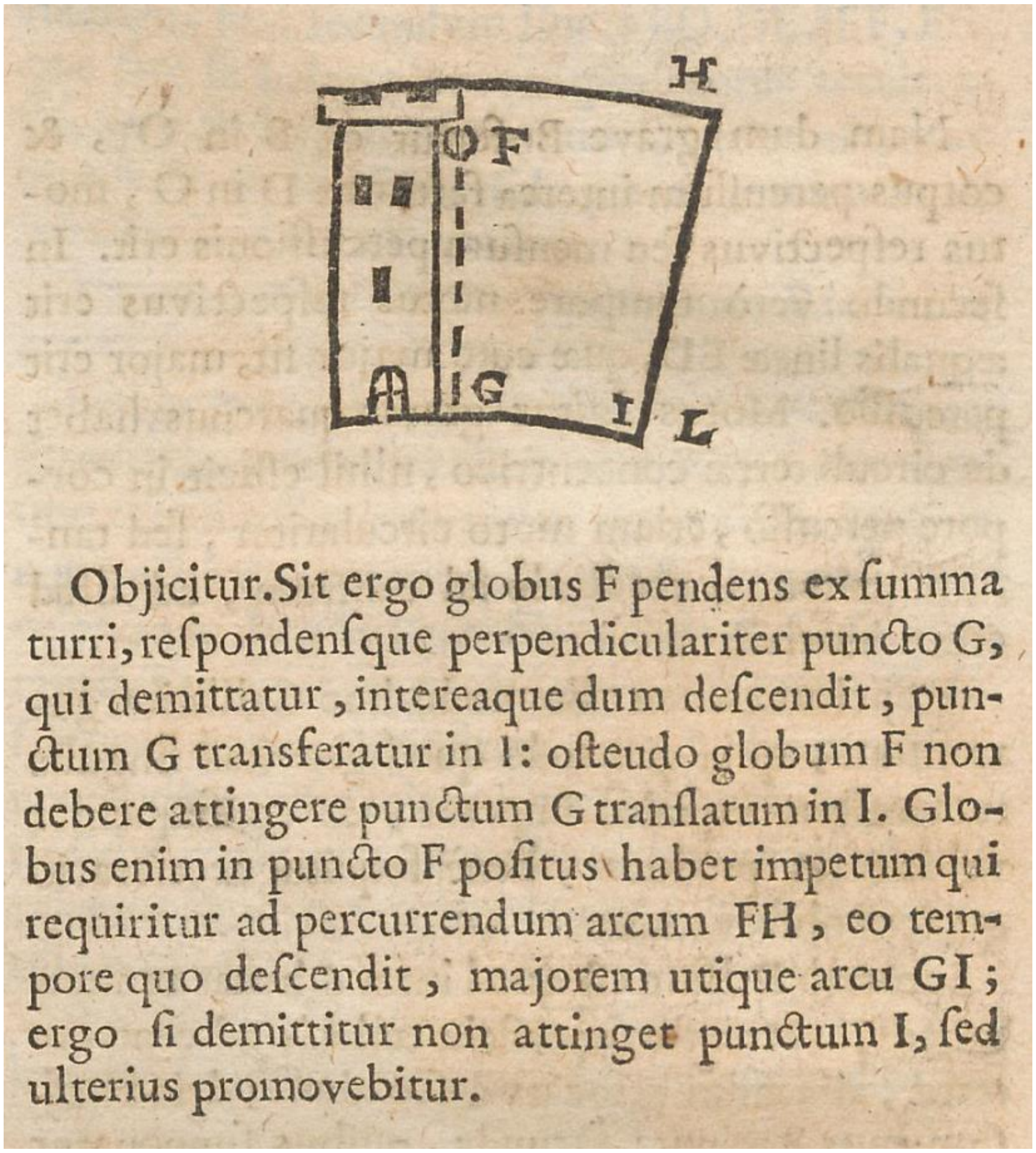

Objicitur. Sit ergo globus F pendens ex ſumma turri, reſpondenſque perpendiculariter puncto G, qui demittatur, intereaque dum deſcendit, punctum G transferatur in I: oſteudo globum F non debere attingere punctum G tranſlatum in I. Globus enim in puncto F poſitus habet impetum qui requiritur ad percurrendum arcum FH, eo tempore quo deſcendit, majorem utique arcu GI; ergo ſi demittitur non attinget punctum I, ſed ulterius promovebitur.

**Figure 3:** This figure shows both the tower diagram and the quoted text describing the diagram.

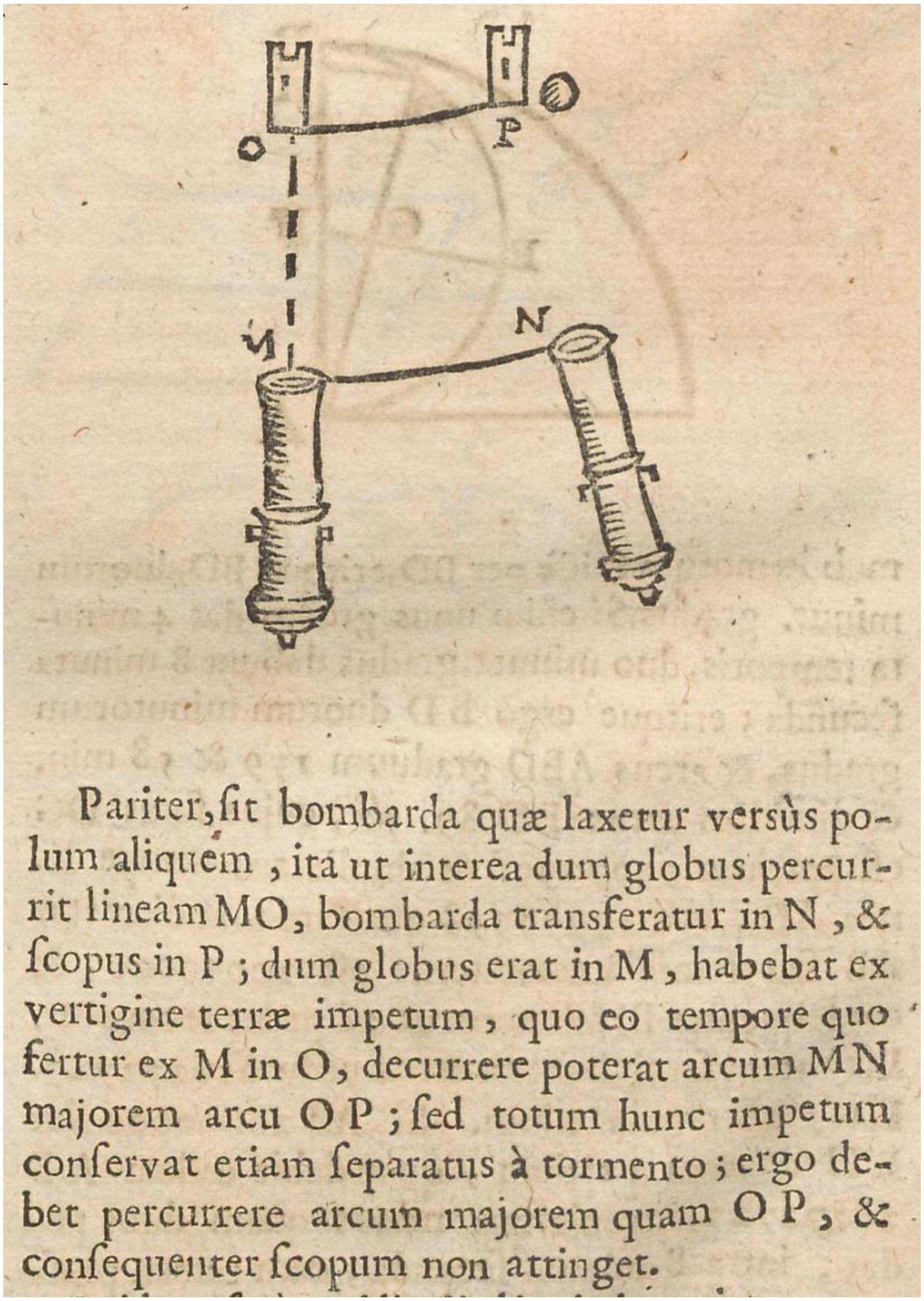

Pariter, ſit bombarda quæ laxetur versùs polum aliquem, ita ut interea dum globus percurrit lineam MO, bombarda transferatur in N, & ſcopus in P; dum globus erat in M, habebat ex vertigine terræ impetum, quo eo tempore quo fertur ex M in O, decurrere poterat arcum MN majorem arcu OP; ſed totum hunc impetum conſervat etiam ſeparatus à tormento; ergo debet percurrere arcum majorem quam OP, & conſequenter ſcopum non attinget.

**Figure 4:** This figure shows both the cannon diagram and the quoted text describing the diagram.

NOTES

[1] G. Coriolis, "Sur les équations du mouvement relatif des systèmes de corps," *Journal de l'Ecole Royale Polytechnique* **15**, 144-154 (1835).

[2] C. M. Graney, "Anatomy of a fall: Giovanni Battista Riccioli and the story of *g*," *Physics Today* **65**(9), 36-40 (2012); "Coriolis effect, two centuries before Coriolis," *Physics Today* **64**(8), 8-9 (2011); "The Coriolis Effect Apparently Described in Giovanni Battista Riccioli's Arguments Against the Motion of the Earth: An English Rendition of *Almagestum Novum* Part II, Book 9, Section 4, Chapter 21, Pages 425, 426-7," arXiv:1012.3642 (2010).

[3] C. F. M. Dechales, *Cursus seu Mundus Mathematicus* (1674), Vol. 3: 297-300, or for the 1690 edition, Vol. 4: 328-330. Translation by C. Graney. Figures are from the 1690 edition (Vol. 4: 328).

[4] The concept described by the Latin word "impetus" could bear much similarity to the modern idea of momentum, depending on the writer using it. See C. M. Graney, "Mass, Speed, Direction: John Buridan's 14th-Century Concept of Momentum," *The Physics Teacher* **51**, 411 and arXiv:1309.4474 (2013); also see C. M. Graney, *Setting Aside All Authority: Giovanni Battista Riccioli and the Science Against Copernicus in the Age of Galileo* (Notre Dame, Indiana: University of Notre Dame Press, 2015), 15-20, 37-43.

[5] Indeed, Isaac Newton and Robert Hooke tried and failed to detect the deflection of a falling body in 1679-1680, after Hooke, at least, had read Riccioli's arguments. Such deflection would elude detection into the 19th century. See *Setting Aside All Authority* (cited in previous note), 115-128.

[6] Two quotations, both from books that present science to a broad audience, illustrate this. First, from Erik Gregersen (ed.), *The Britannica Guide to Heat, Force, and Motion* (New York: Britannica Educational Publishing, 2011), 121:

> Imagine that, somewhere in the Northern Hemisphere, a projectile is fired due south. As viewed from inertial space, the projectile initially has an eastward component of velocity as well as a southward component because the gun that fired it, which is stationary on the surface of Earth, was moving eastward with Earth's rotation at the instant it was fired. However, since it was fired to the south, it lands at a slightly lower latitude, closer to the Equator. As one moves south, toward the Equator, the tangential speed of Earth's surface due to its rotation increases because the surface is farther from the axis of rotation. Thus, although the projectile has an eastward component of velocity (in inertial space), it lands at a place where the surface of Earth has a larger eastward component of velocity. Thus, to the observer on Earth, the projectile seems to curve slightly to the west.... If the projectile were fired to the north, it would seem to curve eastward.

Then, from Philip C. Plait, *Bad Astronomy: Misconceptions and Misuses Revealed, from Astrology to the Moon Landing "Hoax"* (New York: Wiley & Sons, Inc., 2002), 23:

> As you move north from the equator, you can see that your eastward velocity decreases. At the equator you are moving nearly 1,670 kilometers per hour (1,030 miles per hour) to the east.... At Sarasota, Florida… you are moving east at 1,500 kph (930 mph).... Now imagine someone on the equator due south of your [Sarasota] position takes a baseball and throws it directly north, right toward you. As it moves northward, its velocity *eastward* increases relative to the ground. Relative to you, that baseball is moving 1,670 kph – 1,500 kph = 170 kph (1,030 mph – 930 mph = 100 mph) or so to the east by the time it reaches you. Even though the fastball is aimed right at you, it will miss you by a pretty wide margin! By the time it gets to your latitude, it will be a long way to the east of you. That's why cannonballs are deflected as they travel north or south. When they are first shot from the cannon, they have some initial velocity to the east. But if they are fired north, they reach their target moving faster to the east than the ground beneath them.... The reverse is true if it is fired south....

[7] "The Coriolis Effect Apparently Described" (cited in note 2), 11.